# Prefrontal control of the amygdala during real-time fMRI neurofeedback training of emotion regulation

Vadim Zotev [a,*], Raquel Phillips [a], Kymberly D. Young [a], Wayne C. Drevets [a,b], Jerzy Bodurka [a,c,*]

[a] Laureate Institute for Brain Research, Tulsa, OK, USA; [b] Janssen Pharmaceuticals, LLC, of Johnson & Johnson, Inc., Titusville, NJ, USA; [c] College of Engineering, University of Oklahoma, Norman, OK, USA

**Abstract**

We observed in a previous study (PLoS ONE 6:e24522) that the self-regulation of amygdala activity via real-time fMRI neurofeedback (rtfMRI-nf) with positive emotion induction was associated, in healthy participants, with an enhancement in the functional connectivity between the left amygdala (LA) and six regions of the prefrontal cortex. These regions included the left rostral anterior cingulate cortex (rACC), bilateral dorsomedial prefrontal cortex (DMPFC), bilateral superior frontal gyrus (SFG), and right medial frontopolar cortex (MFPC). Together with the LA, these six prefrontal regions thus formed the functional neuroanatomical network engaged during the rtfMRI-nf procedure. Here we perform a structural vector autoregression (SVAR) analysis of the effective connectivity for this network. The SVAR analysis demonstrates that the left rACC plays an important role during the rtfMRI-nf training, modulating the LA and the other network regions. According to the analysis, the rtfMRI-nf training leads to a significant enhancement in the time-lagged effect of the left rACC on the LA, potentially consistent with the ipsilateral distribution of the monosynaptic projections between these regions. The training is also accompanied by significant increases in the instantaneous (contemporaneous) effects of the left rACC on four other regions – the bilateral DMPFC, the right MFPC, and the left SFG. The instantaneous effects of the LA on the bilateral DMPFC are also significantly enhanced. Our results are consistent with a broad literature supporting the role of the rACC in emotion processing and regulation. Our analysis provides, for the first time, insights into the causal relationships within the network of regions engaged during the rtfMRI-nf procedure targeting the amygdala. It suggests that the rACC may constitute a promising target for rtfMRI-nf training along with the amygdala in patients with affective disorders, particularly posttraumatic stress disorder (PTSD).

*Keywords:* Neurofeedback, real-time fMRI, emotion regulation, amygdala, rostral anterior cingulate cortex (rACC), effective connectivity, vector autoregression (VAR), structural vector autoregression (SVAR)

## Introduction

Interactions between various regions of the prefrontal cortex (PFC) and the amygdala play a fundamental role in processing and regulation of human emotions. One widely accepted neural model of emotion regulation [1] draws a distinction between voluntary and automatic regulation processes and delineates neural systems involved in each type of emotion regulation. The model posits that dorsal prefrontal cortical regions, including bilateral dorsolateral prefrontal cortex (DLPFC), bilateral dorsomedial prefrontal cortex (DMPFC), and bilateral dorsal anterior cingulate cortex (ACC), are involved in different subprocesses associated with voluntary emotion regulation [1]. Neural processing within these regions may be modulated by the ventromedial prefrontal cortex and the orbitofrontal cortex (OFC) regions, both having direct and extensive connections with the amygdala [1]. The model further suggests that left rostral (pregenual) ACC (rACC), bilateral subgenual ACC, bilateral OFC, bilateral DMPFC, and midline dorsal ACC are implicated (with contributions from the hippocampus and parahippocampus) in various subprocesses associated with automatic emotion regulation [1].

Functional neuroimaging studies of voluntary emotion regulation generally provide an explicit instruction to regulate emotion and a cognitive strategy to achieve such regulation. Typical regulation methods include reappraisal [2-8], i.e. a cognitive re-interpretation of emotionally evocative stimuli, and suppression [9-11], i.e. a voluntary inhibition of reaction to emotional stimuli. Blood-oxygenation-level-dependent (BOLD) fMRI studies involving reappraisal of negative emotional experiences have demonstrated negative (inverse) functional coupling between the PFC and the amygdala, such that increased activity of PFC regions during reappraisal is associated with a reduction in the amygdala BOLD response to disturbing or aversive stimuli, and also with a reduction in the intensity of the nega-tive affect [2,5,7]. This functional interaction putatively represents the top-down inhibitory control of the amygdala by the PFC. Negative functional coupling also was observed in a psychophysiological interaction (PPI) [12] analysis, exploring functional connectivity between the amygdala and the PFC specific to the reappraisal task [8]. (An earlier work, however, suggested that the coupling was positive [6], potentially due to the methodological differences with [8]). Consistent with the model [1], most studies of voluntary emotion regulation have shown activity of the dorsal ACC, though involvement of the rACC also has been reported [8,9].

*Corresponding authors. E-mail: vzotev@laureateinstitute.org (V. Zotev), jbodurka@laureateinstitute.org (J. Bodurka)

In contrast, neuroimaging studies of automatic emotion regulation commonly present emotionally evocative stimuli as task-irrelevant emotional distracters during an ongoing main task, as exemplified by the emotional Stroop task and its modifications [13-19]. fMRI studies utilizing such tasks have consistently shown greater BOLD activity in the left rACC for emotional than for neutral stimuli [13-16]. Recent studies of emotional conflict [17-19] demonstrated that, while emotional conflict monitoring was associated wth activity of the DMPFC and DLPFC, emotional conflict resolution more specifically was related to activity of the left rACC. PPI analysis revealed negative functional coupling between the left rACC and the amygdala for the emotional conflict resolution, such that increases in BOLD activity of the left rACC were accompanied by reductions in amygdala activity induced by the emotional conflict. The apparent top-down inhibitory effect of the left rACC on the amygdala suggested by these associations was supported using dynamic causal modeling (DCM) [20] of the interactions between the two regions [17]. These results are consistent with the model [1], pointing to the important role of the left rACC in automatic emotion regulation. It has been suggested that rACC may contribute to both appraisal/expression and regulation of emotion (Fig. 3 in [21]).

Recently, we demonstrated that healthy volunteers could learn to self-regulate BOLD activity in their left amygdala (LA) using real-time fMRI neurofeedback (rtfMRI-nf) [22]. During the rtfMRI-nf procedure, the participants were asked to induce positive emotions by evoking happy autobiographical memories, while simultaneously trying to control and raise the neurofeedback bar on the screen. The height of the bar represented BOLD activity in the LA region of interest (ROI). Importantly, the target level for the neurofeedback bar was raised from run to run in a linear fashion. In the group analysis, the LA BOLD activity exhibited a significant increase (positive linear trend) across the neurofeedback training runs [22]. Moreover, six other brain regions showed a significant enhancement (positive linear trend) in their functional connectivity with the LA as the rtfMRI-nf training progressed. These regions were located near the medial wall of the PFC, and included the left rACC, bilateral DMPFC, bilateral superior frontal gyrus (SFG) and right medial fronto-topolar cortex (MFPC) [22]. Functional neuroimaging studies have consistently shown involvement of medial PFC regions in internally focused emotion processing [3].

Despite successful proof-of-concept applications of rtfMRI-nf [23] for self-regulation of various brain regions and networks relevant to emotion processing, (e.g. [22,24-31]), however, the neural mechanisms underlying the neurofeedback training effect, and the specific nature of the interactions among the engaged brain regions remain unclear. Functional connectivity analyses provide information about temporal correlations of BOLD fMRI activities in various brain areas, but do not yield insights into causal relationships among them. Thus, studies of effective connectivity of brain regions engaged during rtfMRI-nf training are needed, including the experimental paradigm described above [22]. Real-time measures of effective connectivity can also be used to provide connectivity-based rtfMRI-nf [32]. Furthermore, involvement of different subprocesses of voluntary and automatic emotion regulation [1] during rtfMRI-nf training requires careful evaluation. On the one hand, the emotion self-induction with rtfMRI-nf, employed in [22], constitutes voluntary emotion regulation. On the other hand, the two experimental tasks – inducing positive emotion and controlling the neurofeedback bar on the screen – provide mutual interference, and success of the rtfMRI-nf training depends on a participant's ability to achieve proper balance between the two tasks while performing them simultaneously in real time. In this respect, the rtfMRI-nf training of emotional self-regulation exhibits some parallels with the experimental paradigms used to study automatic emotion regulation.

In this work, we report an analysis of effective connectivity for the system of regions showing enhanced functional connectivity with the left amygdala during the rtfMRI-nf training [22]. The analysis is based on structural vector autoregression (SVAR), a promising method for effective connectivity modeling [33]. The purpose of this analysis is to elucidate interactions between the amygdala and the PFC, which are specific to the rtfMRI-nf procedure [22]. Understanding these interactions may conceivably lead to the development of novel rtfMRI-nf paradigms for training of emotional self-regulation, including paradigms that provide new therapeutic approaches for individuals suffering from mood and anxiety disorders.

**Methods**

*Subjects and Procedure*

The study was conducted at the Laureate Institute for Brain Research. The research protocol was approved by the University of Oklahoma Institutional Review Board. Human research in this study was conducted according to the principles expressed in the Declaration of Helsinki.

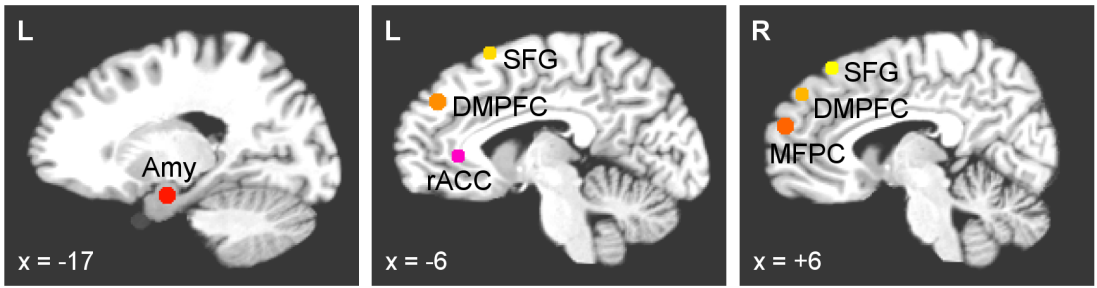

**Figure 1. Regions of interest for the effective connectivity analysis.** Six brain regions exhibited a significant enhancement in functional connectivity with the left amygdala during the rtfMRI neurofeedback training with positive emotion induction [22]. They included: the left rostral anterior cingulate cortex (rACC, BA 24), bilateral dorsomedial prefrontal cortex (DMPFC, BA 9), bilateral superior frontal gyrus (SFG, BA 6,8), and right medial frontopolar cortex (MFPC, BA 10). The 10 mm diameter regions of interest (ROIs) in those areas are projected onto the standard anatomical template (TT_N27) in the stereotaxic array of Talairach and Tournoux [34].

All subjects gave written informed consent to participate in the study and received financial compensation. Twenty eight healthy male volunteers (age 28±9 years) participated in the rtfMRI-nf study described in detail in [22]. The participants were randomly assigned to either the experimental group (EG, 14 subjects) or the control (sham) group (CG, 14 subjects). During the experiment, each participant was asked to perform a positive emotion induction task based on retrieval of happy autobiographical memories, while simultaneously trying to raise the rtfMRI-nf bar on the screen [22]. The subjects in EG were provided with rtfMRI-nf based on BOLD activity in the LA ROI. The center of this 14 mm diameter ROI was selected at the locus: $x=-21$, $y=-5$, $z=-16$, in the stereotaxic array of Talairach and Tournoux [34] based on a meta-analysis of functional neuroimaging studies investigating the role of the amygdala in emotion processing [35]. The subjects in CG received sham rtfMRI-nf based on BOLD activity in the left horizontal segment of the intraparietal sulcus (HIPS) ROI. This ROI was centered at the locus: $x=-42$, $y=-48$, $z=48$, taken from a review of fMRI studies investigating the role of HIPS in number processing [36]. Thus, the sham neurofeedback was based on BOLD activity within a region presumably *not* involved in emotion regulation.

The rtfMRI-nf experiment included six fMRI runs each lasting 8 min 40 s: Rest, Practice, Run 1, Run 2, Run 3, and Transfer [22] (abbreviated as RE, PR, R1, R2, R3, and TR, respectively). Each run (except Rest) consisted of alternating blocks of Rest, Happy Memories, and Count conditions [22]. The condition blocks were 40 s long for Run 1, Run 2, Run 3, and the Transfer run. Each Happy Memories condition block was preceded by a Rest block and followed by a Count block. Instructions for each condition were provided to a subject inside an MRI scanner as visual cues via the neurofeedback GUI screen [22]. For the Rest conditions, the participants were instructed to rest while viewing the screen. For the Count conditions, the subjects were asked to count backwards from 100 by subtracting a given integer. For the Happy Memories conditions during the neurofeedback runs (Practice, Run 1, Run 2, and Run 3), the participants were instructed to feel happy by evoking and contemplating happy autobiographical memories, while also trying to control and raise the neurofeedback bar on the screen. The bar height was updated every 2 s. The target level for the neurofeedback bar was raised in equal increments from run to run. For the Happy Memories conditions during the Transfer run, no neurofeedback was provided, but the subjects were asked to feel happy using the same strategies as during the rtfMRI-nf training. Details of the experimental protocol and instructions given to the participants can be found in our previous work [22].

All functional and structural MR images were acquired using a General Electric Discovery MR750 3T MRI scanner with a standard 8-channel receive head coil array as described in [22]. A gradient-recalled echo-planar imaging (EPI) sequence with sensitivity encoding (SENSE) [37] and two-fold acceleration ($R$=2) was employed for fMRI. The sequence provided the whole-brain coverage with $1.875 \times 1.875 \times 2.9$ mm$^3$ spatial resolution and temporal resolution equal to the fMRI repetition time $TR$=2000 ms. A T1-weighted magnetization-prepared rapid gradient-echo (MPRAGE) sequence with SENSE $R$=2 was used to acquire anatomical brain images with $0.9375 \times 0.9375 \times 1.2$ mm$^3$ spatial resolution.

*Regions of Interest*

The seven ROIs for the network analysis were selected based on the functional connectivity results reported in our previous study [22]. The ROIs are shown in Figure 1. Each ROI was defined as a sphere 10 mm in diameter and positioned as follows. The ROI in the LA region was centered at the locus $(-17, -7, -16)$ that exhibited the largest difference in mean BOLD activity levels (within the LA region) between the EG and CG groups for the Happy Memories conditions during both Run 3 and the Transfer run. The other six ROIs showed a significant enhancement in functional connectivity strength with this LA seed ROI for EG. This connectivity enhancement (positive linear trend) was statistically significant both across the neurofeedback training runs (RE, PR, R1, R2, R3) and across the entire experiment includ-ing the Transfer run (RE, PR, R1, R2, R3, TR), as described in detail in [22]. The six ROIs were located in the following brain areas and centered at the following points based upon our previously reported results ([22]): the left rACC (BA 24) at (−3, 34, 5); the left DMPFC (BA 9) at (−6, 45, 34); the right DMPFC (BA 9) at (3, 47, 38); the right MFPC (BA 10) at (5, 56, 21); the left SFG (BA 6) at (−9, 17, 62); and the right SFG (BA 8) at (9, 31, 54). While many brain regions exhibited functional connectivity with the LA during the experiment [22], the significant enhancement in the connectivity strength for these six regions indicated their special role during the rtfMRI-nf training. For convenience, we refer to the seven regions in Fig. 1 as a "network", with understanding that these regions may *potentially* form a network or belong to a broader emotion regulation network.

*Network Modeling*

We performed analyses of effective connectivity for the network in Fig. 1 using the structural vector autoregression (SVAR) method described in [33]. SVAR combines the capabilities of the structural equation modeling (SEM), which is a hypothesis-driven approach, and the vector autoregression (VAR, Granger causality [38]), which is a data-driven approach. SVAR can model both instantaneous (contemporaneous) and lagged effects among network regions using a unified analytical framework. While no interactions within the brain are truly instantaneous, the inclusion of the instantaneous effect terms makes it possible to model interactions with delay times much shorter than the lag time set by the temporal resolution of fMRI.

A multivariate SVAR model of the first order (number of lags $p=1$) for a network of $n$ ROIs is defined as follows [33]:

SVAR(1):

$$X(t) = A_0 X(t) + A_1 X(t-1) + b_1 z_1(t) + \ldots + b_m z_m(t) + e(t) \quad (1)$$

$$X(t) = \begin{bmatrix} x_1(t) \\ \vdots \\ x_n(t) \end{bmatrix}, A = \begin{bmatrix} \alpha_{11} & \cdots & \alpha_{1n} \\ \vdots & \ddots & \vdots \\ \alpha_{n1} & \cdots & \alpha_{nn} \end{bmatrix}, b = \begin{bmatrix} \beta_1 \\ \vdots \\ \beta_n \end{bmatrix}, e(t) = \begin{bmatrix} \varepsilon_1(t) \\ \vdots \\ \varepsilon_n(t) \end{bmatrix}$$

Here, $X(t)$ is a vector consisting of fMRI signal values $x_i(t)$ for $n$ ROIs at time point $t$, and $X(t-1)$ is a vector of fMRI signals for the same ROIs at the preceding time point $t-1$. The time points correspond to consecutive fMRI volumes, and the minimum nonzero lag time is equal to the fMRI repetition time *TR*. The $n \times n$ matrices $A_0$ and $A_1$ contain path coefficients $\{\alpha_{ij}\}$ for different pairs of ROIs. A path coefficient $\alpha_{ij}$ specifies a directional effect of the $j$-th ROI on the $i$-th ROI. The matrix $A_0$ describes instantaneous effects within the network. The diagonal elements of $A_0$ are zeros, and the maximum number of free parameters is $n(n-1)/2$ according to [33]. The remaining path coefficients have to be fixed to predefined nonzero values or set to zero. Thus, the general structure of $A_0$ must be defined prior to the SVAR analysis based on some hypothesis about the instantaneous effects among the network regions. The matrix $A_1$ describes lagged effects with lag 1 (i.e. *TR*) within the network. No a priori assumptions about properties of $A_1$ are needed, so determination of path coefficients for the lagged effects is data-driven. The functions $z_1(t)\ldots z_m(t)$ in Eq (1) are exogenous variables, such as physiological confounds or experimental design parameters, which are independent of the interactions within the network. Their effects are described in the model by vectors $b_k$, $k=1\ldots m$. The $e(t)$ is a vector of residuals $\{\varepsilon_i(t)\}$, assumed to be serially and mutually independent with Gaussian distributions [33].

A first-order multivariate VAR model ($p=1$) for the same network is defined as follows:

VAR(1):

$$X(t) = A_1 X(t-1) + b_1 z_1(t) + \ldots + b_m z_m(t) + e(t) \quad (2)$$

It can be considered a particular case of the first-order SVAR model described by Eq (1). In VAR, the lagged effects are modeled explicitly by elements of the $n \times n$ matrix $A_1$, and require no prior assumptions. The instantaneous effects are accounted for by the residuals in the vector $e(t)$. However, the residuals in this case can no longer be assumed to be serially and mutually independent [33].

*Data Analysis*

The fMRI data processing and analysis were performed using Analysis of Functional NeuroImages (AFNI) software [39,40]. The AFNI program 1dSVAR.R was used for multivariate SVAR analysis, Eq (1), and the program 1dGC.R was employed for multivariate VAR, Eq (2). The programs are distributed with AFNI and described in [33,41]. They were customized for the analyses in the present study. Analysis of percent BOLD signal changes for the seven ROIs in Fig. 1 also was performed in AFNI using the general linear model (GLM) framework, as described in [22]. Statistical data analyses were conducted using Statistical Package for Social Sciences (IBM SPSS Statistics 20).

Pre-processing of single-subject fMRI data for the subsequent network analysis included correction of cardiorespiratory artifacts using RETROICOR [42], slice timing correction, and volume registration. The seven ROIs, defined in the Talairach space (Fig. 1), were transformed to an individual subject's EPI image space using a high-resolution structural brain image for that subject, acquired prior to the fMRI experiment. Four additional ROIs were defined bilaterally (on the left and on the right, to avoid asymmetry) within white matter and ventricle CSF, and were similarly transformed. Each ROI in the EPI space contained approximately 50 voxels. The EPI images were spatially smoothed using a Gaussian kernel with full-width at half-maximum (FWHM) of 5 mm. No temporal filtering or baseline correction was applied to the data prior to the network analyses. Time courses of the mean fMRI signals for the selected ROIs were exported and used as input time series for the network modeling.

The multivariate SVAR and VAR analyses were performed according to Eqs (1) and (2) for the network of seven ROIs ($n=7$) in Fig. 1 for each of the six experimental runs. Following [41], we selected the first-order model for the lagged effects. The exogenous variables $\{z_k(t)\}$ included six fMRI motion parameters, time courses of the four ROIs within white matter and ventricle CSF, and six Legendre polynomials for modeling the baseline. Thus, the total number of covariates in Eqs (1) and (2) for $n=7$ was $m=16$.

The SVAR modeling, Eq (1), is more challenging than the VAR modeling, Eq (2), because it requires a priori assumptions about the structure of the matrix $A_0$ describing instantaneous effects. For $n=7$, there are 42 possible directional effects among different regions, but no more than 21 elements in the matrix $A_0$ can be optimized simultaneously (see *Network Modeling*). This makes the number of possible structural models, that should be optimized and compared, prohibitively large. In the present work, however, we were primarily interested in those interactions that exhibited significant progressive changes with the rtfMRI-nf training. This consideration provided an additional criterion, which we used to simplify the SVAR model. This criterion was applied as follows. A "star" model for instantaneous effects was defined and optimized for each of the seven ROIs. Each star model only described directional effects of a selected region onto the other six regions, so the matrix $A_0$ had only six free parameters in each case (see *SVAR Analysis*). Upon examination of the SVAR results for the seven star models, we selected three ROIs that showed the most significant linear trends in their instantaneous interactions across the rtfMRI-nf training runs (see *SVAR Analysis* and *Discussion* for details).

For the chosen system of three ROIs ($n=3$), we defined and optimized a total of 24 SVAR models. These models included four additional censor covariates (yielding $m=20$), each equal to 1 for one of the 40-s long Count condition blocks (as defined for Runs 1-3 and the Transfer run in [22]), and 0 for all other points. Such censoring effectively excluded the Count condition blocks from the analysis. A SVAR model for three ROIs allows simultaneous optimization of as many as three path coefficients for instantaneous effects (see *Network Modeling*). However, comparison of different structural models using the $\chi^2$ criterion is only possible if less than three (for $n=3$) model parameters are optimized at the same time ($df>0$). Therefore, the SVAR modeling of the system with three ROIs was performed in two steps. *First*, all possible models with two instantaneous effects were optimized. The matrix $A_0$ in each case had two free parameters ($df=1$), and all the other matrix elements were set to zero. Twelve structural models were optimized in this way. While there are 15 pairs of directional effects for a system of three ROIs, structural models with non-recursive paths (i.e. A=>B & B=>A, three in this case) are numerically unstable in SEM analysis. *Second*, for each of the 12 models with optimized parameters, two nested structural models were defined by inclusion of a third instantaneous interaction with one of two possible directions. For example, if the instantaneous effects A=>B and A=>C (with A,B, and C denoting the three regions) were optimized at the first step, one nested model was defined with B=>C interaction, and the other – with C=>B interaction. The matrix $A_0$ in each case had one free parameter ($df=2$) and two constant elements from the previous step, with the remaining elements set to zero. Thus, 24 models for instantaneous effects were defined, optimized, and compared using the $\chi^2$ measure of fit quality.

Each single-subject analysis by means of 1dGC.R or 1dSVAR.R programs yielded values of path coefficients $\{\alpha_{ij}\}$ together with corresponding *t*-statistics. Group analyses of the results were performed using the same programs. Each group analysis was a meta-analysis utilizing both path coefficients $\{\alpha_{ij}\}$ and their respective *t*-values for each subject in a group. The analysis provided estimates of a group path coefficient and its statistical significance (*P*-value, two-tailed, uncorrected) for each interaction within the network. Correction for multiple comparisons was performed using the false discovery rate (FDR) procedure [43], implemented in 3dFDR AFNI program. This program was applied to a column of uncorrected *P*-values. Trends in group effects across experimental runs were evaluated using the GLM for Repeated Measures analysis in SPSS, applied to path coefficient values $\{\alpha_{ij}\}$ (without *t*-statistics) for multiple runs for all subjects in a given group. Similar trend analyses were conducted for percent BOLD signal change results for each of the seven ROIs.

**Results**

*ROI Analysis*

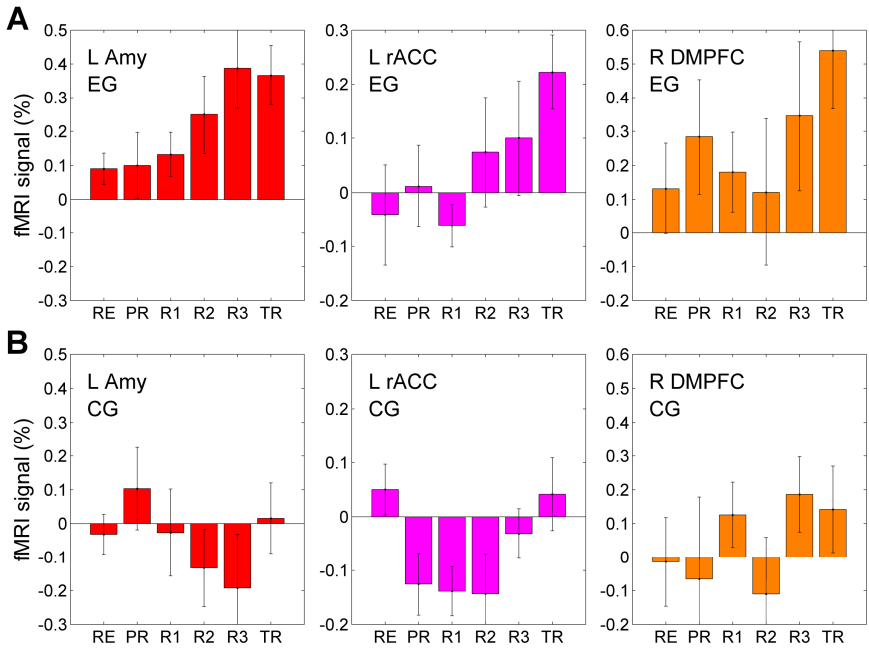

**Figure 2. Learned enhancement of control over BOLD activity and emotion induction.** (A) Mean BOLD signal activity of the left amygdala during the rtfMRI neurofeedback (rtfMRI-nf) training for the experimental group (EG). The EG subjects received rtfMRI-nf based on the BOLD activity in the left amygdala ROI. Each bar represents a group average (mean±sem) of percent BOLD signal changes for the Happy Memories condition vs Rest condition for each of the six experimental runs: Rest (RE), Practice (PR), Run 1 (R1), Run 2 (R2), Run 3 (R3), and Transfer (TR). The enhancement in the left amygdala activity (red) was accompanied by increased activities of the left rACC (magenta), the right DMPFC (orange), as well as the other ROIs depicted in Fig. 1. (B) Lack of learned control over BOLD activity of the left amygdala and other regions for the control (sham) group (CG). The CG subjects received sham rtfMRI-nf based on BOLD activity in the left horizontal segment of the intra-parietal sulcus (HIPS), presumably *not* involved in emotion regulation.

Figure 2 illustrates BOLD activity properties for three representative ROIs in the network (Fig. 1) – the LA, the left rACC, and the right DMPFC – for the six experimental runs. The results in Fig. 2A correspond to EG, and the results in Fig. 2B – to CG. Each bar in the figures represents a mean percent BOLD signal change for a given ROI, averaged for Happy Memories conditions during a given run and across all subjects in a given group. The mean ROI results for each participant were obtained from the GLM analysis described in [22]. The error bars are standard errors of the means (sem) across the subjects. The results for the LA ROI in Fig. 2 differ slightly from those reported in our previous study [22], because they correspond to the LA seed ROI defined based on the functional contrast between EG and CG (see *Regions of Interest*) rather than the LA target ROI based on the published meta-analysis (see *Subjects and Procedure*). The abbreviation "LT" in the text below refers to a linear trend, and $t(13)$ is the linear trend $t$-statistics (for 14 subjects), corresponding to $F(1,13)$ trend statistics in the GLM for Repeated Measures analysis in SPSS (see *Data Analysis*).

The LA BOLD activity for EG (Fig. 2A) exhibited a significant positive linear trend across the neurofeedback training runs with the Rest run as the starting point (LT(RE…R3): $t(13)=2.467$, $P<0.028$) and across the entire experiment including the Transfer run (LT(RE…TR): $t(13)=3.170$, $P<0.007$). The mean BOLD activity levels during the Transfer run and Run 3 did not differ significantly from each other (TR vs R3: $t(13)=-0.195$, $P<0.849$). For the CG results in Fig. 2B, there was no significant linear trend for the LA BOLD activity across the experiment (LT(RE…TR): $t(13)=0.691$, $P<0.502$). Comparison of the mean BOLD activity levels between the EG and CG groups showed significant differences for Run 2 ($t(26)=2.360$, $P<0.026$), Run 3 ($t(26)=2.887$, $P<0.008$), and the Transfer run ($t(26)=2.556$, $P<0.017$).

The left rACC results for EG (Fig. 2A) showed a linear trend across the experiment that was nonsignificant but trended toward significance (LT(RE…TR): $t(13)=2.013$, $P<0.065$). The mean activity levels for the Transfer run and Run 3 did not exhibit a significant difference (TR vs R3: $t(13)=1.063$, $P<0.307$). For the CG results in Fig. 2B, there was no significant linear trend for the rACC BOLD activity levels across the experiment (LT(RE…TR): $t(13)=0.505$, $P<0.622$). Comparison of the mean activity levels between EG and CG showed trends toward differences for Run 2 ($t(26)=1.735$, $P<0.095$) and for the Transfer run ($t(26)=1.875$, $P<0.072$).

The right DMPFC BOLD activity for EG (Fig. 2A) exhibited a linear trend across the experiment that was nonsignificant with a trend toward significance (LT(RE…TR): $t(13)=2.085$, $P<0.057$). There was no significant difference between the mean activity levels for the Transfer run and Run 3 (TR vs R3: $t(13)=0.598$, $P<0.560$). For the CG results in Fig. 2B, there was no significant linear trend for the right DMPFC BOLD activity levels across the experiment (LT(RE…TR): $t(13)=1.033$, $P<0.320$). Comparison of the mean BOLD activity levels between EG and CG showed a trend toward a difference for the Transfer run ($t(26)=1.839$, $P<0.077$).

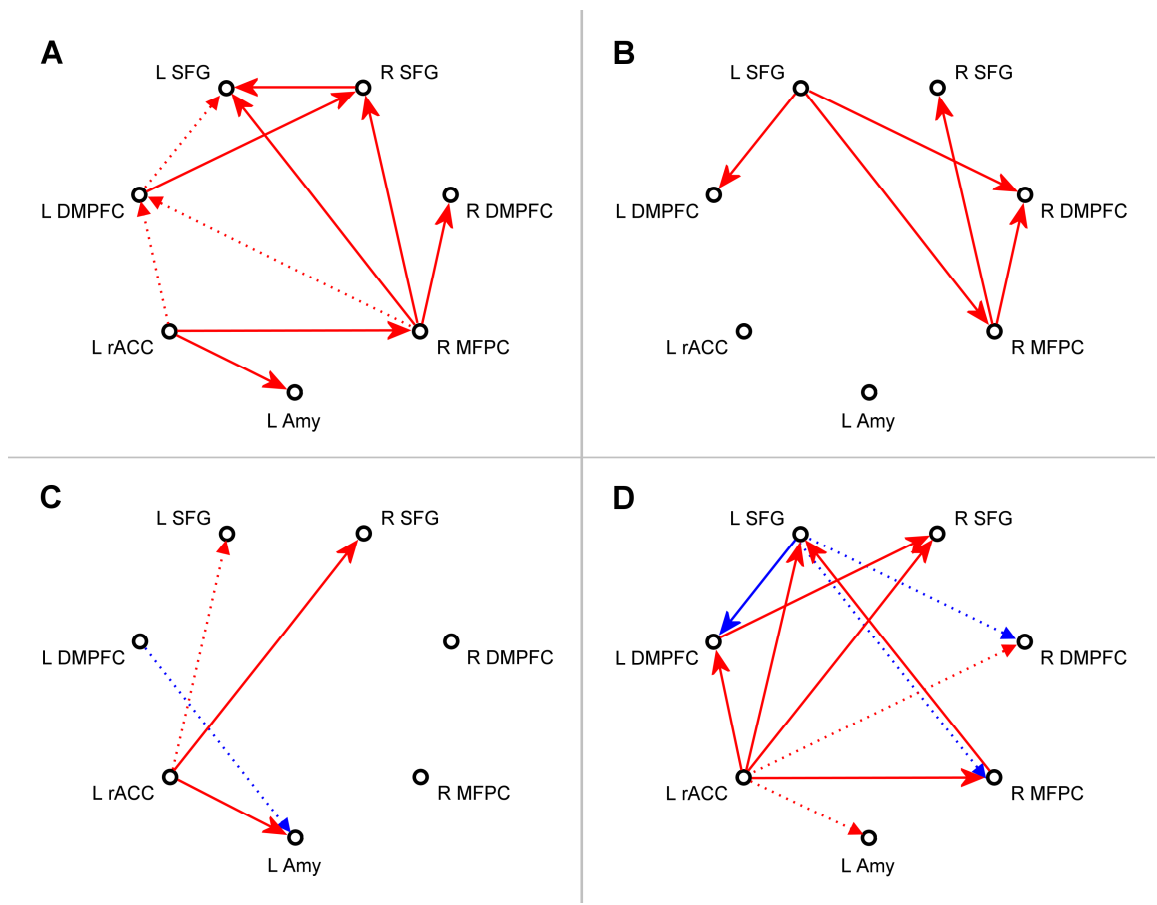

**Figure 3. Interactions within the network suggested by the multivariate VAR analysis.** Results of the multivariate first-order vector autoregression (VAR) analysis for the network of seven ROIs depicted in Fig. 1. The four subplots show meta-analytic group statistics for path coefficients for the following groups and contrasts. (A) Experimental group (EG), neurofeedback Run 3. (B) Control group (CG), Run 3. (C) Difference between Run 3 and Rest for EG. (D) Difference between Run 3 for EG and Run 3 for CG. Red arrows denote augmentation effects (path coefficient α>0), and blue arrows – inhibition effects (path coefficient α<0). In (A) and (B), solid arrows correspond to effects with FDR $q<0.05$, and dotted arrows – to effects with $0.05 \leq q < 0.1$. In (C) and (D), solid arrows correspond to results with uncorrected $P<0.05$, and dotted arrows – to results with $0.05 \leq P < 0.1$.

Correlation analysis for EG revealed positive across-subjects correlations between the mean BOLD activity levels for the left rACC on the one hand, and the mean BOLD activity levels for the LA and the right DMPFC, on the other hand. For example, for the Transfer run: LA vs left rACC: $r=0.544$, $P<0.040$; right DMPFC vs left rACC: $r=0.483$, $P<0.080$.

Results of the VAR and SVAR analyses reported below were similarly tested for linear trends across experimental runs and group differences as signatures of the rtfMRI-nf training effects.

*VAR Analysis*

Results of the multivariate VAR analysis, Eq (2), appear in Figure 3 and Table 1. The figure schematically depicts directional lagged effects (with the lag time of 2 s equal to the repetition time *TR* in the rtfMRI-nf experiment) suggested by the group-level analyses. Table 1 includes group path coefficients for the effects of the left rACC on the other six regions. The results in Fig. 3 and Table 1 were obtained from the group meta-analysis procedure described above (see *Data Analysis*). The one-group results in Fig. 3A and Fig. 3B were corrected for multiple comparisons using the FDR procedure (see *Data Analysis*) and thresholded using FDR $q$-values. The group differences in Fig. 3C and Fig. 3D did not survive the FDR correction, and were thresholded using uncorrected $P$-values. The notation "=>" in the text, figures, and tables denotes a directional effect of one region onto the other.

Figure 3A shows VAR results for the last neurofeedback training run (Run 3) for EG. The results suggest that the left rACC exerted significant effects on the LA (rACC=>LA: $α=0.0857$, $q<0.024$) and the right MFPC (Table 1, column A). VAR results corresponding to Run 3 for CG appear in Fig. 3B. The results revealed no significant effects involving either the LA or the left rACC (Table 1, column B). Figure 3C shows group differences between the VAR results for Run 3 and the Rest run for EG. The results demonstrate that the effects of the left rACC were enhanced during the rtfMRI-nf training for EG. The enhancement was significant for the LA (rACC=>LA: Δα=0.1099, $P<0.010$) and the right SFG (Table 1, column C). Group differences in the VAR results for Run 3 between EG and CG are exhibited in Fig. 3D. The results indicate that the effects of the left rACC on the other six network regions were stronger for EG than for CG. The group differences for the left DMPFC, the right MFPC, the left SFG, and the right SFG were significant, while the group differences for the LA (rACC=>LA: Δα=0.1102, $P<0.077$) and the right DMPFC trended toward significance (Table 1, column D). The left rACC effects for the Transfer run (not shown) for EG exhibited nonsignificant reductions compared to Run 3.

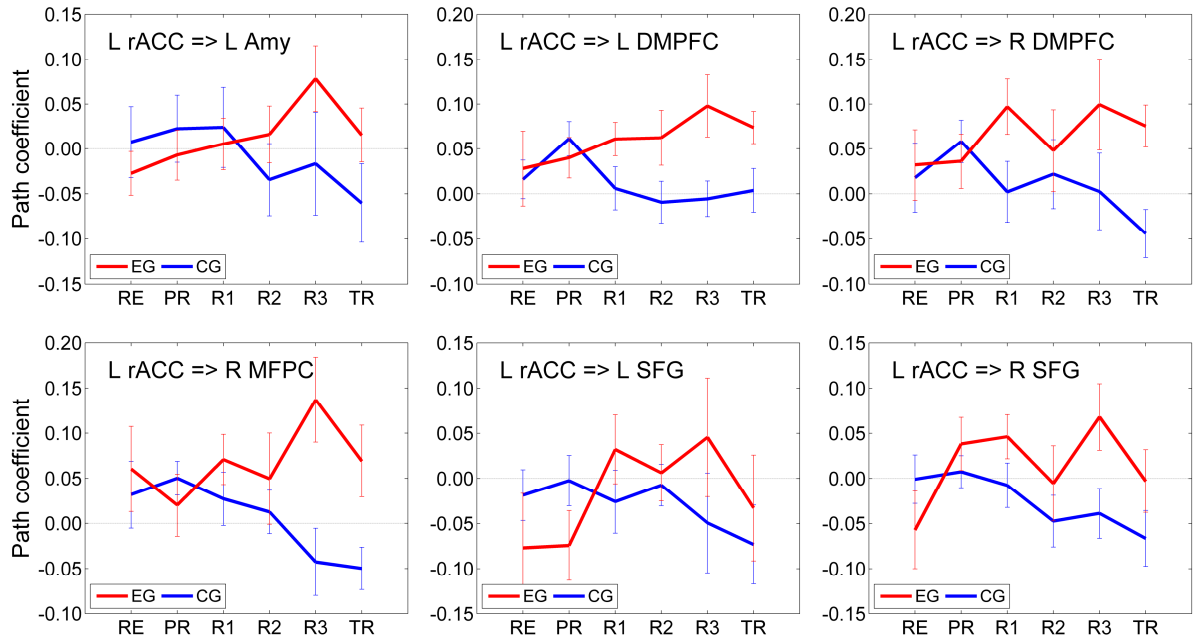

**Figure 4. Effects of the rACC on the other six network regions suggested by the multivariate VAR analysis.** Average path coefficient values (mean±sem) describing the effects of the left rACC on the other six network regions based on the analysis illustrated in Fig. 3. The results for each of the six experimental runs are shown in red for the experimental group (EG) and in blue for the control group (CG).

The average VAR path coefficients (mean±sem) describing the lagged effects of the left rACC on the other six network regions appear in Figure 4, and the corresponding linear trend statistics are included in Table 2. Group-level trends across multiple experimental runs were evaluated using the GLM for Repeated Measures analysis as described above (see *Data Analysis*). Table 2 shows linear trend *t*-statistics with the corresponding group *P*-values for six rACC effects for both EG and CG.

The left rACC effects for EG exhibited significant positive linear trends across the neurofeedback runs (RE, PR, R1, R2, R3) for four regions. These regions included the LA (rACC=>LA: LT(RE…R3), $t(13)=2.348$, $P<0.035$), the right MFPC, the left SFG and the right SFG (Table 2, column EG). The results for the left and right DMPFC showed more significant linear trends across the entire experiment (rACC=>left DMPFC: LT(RE…TR), $t(13)=2.215$, $P<0.045$; rACC=>right DMPFC: LT(RE…TR), $t(13)=2.088$, $P<0.057$). The left rACC effects for CG exhibited negative trends (Table 2, column CG). The negative trends were more significant across the entire experiment (for example, rACC=>right DMPFC: LT(RE…TR), $t(13)=-2.852$, $P<0.014$; rACC=>right MFPC: LT(RE…TR), $t(13)=-3.546$, $P<0.004$).

Beyond the left rACC effects specified in Fig. 4 and Table 2, only four other VAR interactions (out of total 49) exhibited linear trends that were either significant or approaching significance for EG: rACC=>self (LT(RE…R3), $t(13)=2.789$, $P<0.015$); left DMPFC =>right DMPFC (LT(RE…R3), $t(13)=-1.837$, $P<0.089$); left DMPFC=>left SFG (LT(RE…R3), $t(13)=-2.346$, $P<0.036$); and right MFPC=>left SFG (LT(RE…R3), $t(13)=2.083$, $P<0.058$). Therefore, the left rACC had more VAR effects that showed significant trends across the neurofeedback runs than the other six regions combined. This result is consistent with the group difference statistics in Fig. 3C and Fig. 3D. Because such linear trends were associated mainly with the left rACC, the trend statistics in Table 2 did not survive FDR correction for multiple comparisons over the 49 VAR interactions. However, when FDR correction was applied to the linear trend statistics for the six rACC effects only (Fig. 4, Table 2), as the most relevant ones, the corrected results approached significance, with FDR $q<0.055$ for the four most significant trends in Table 2.

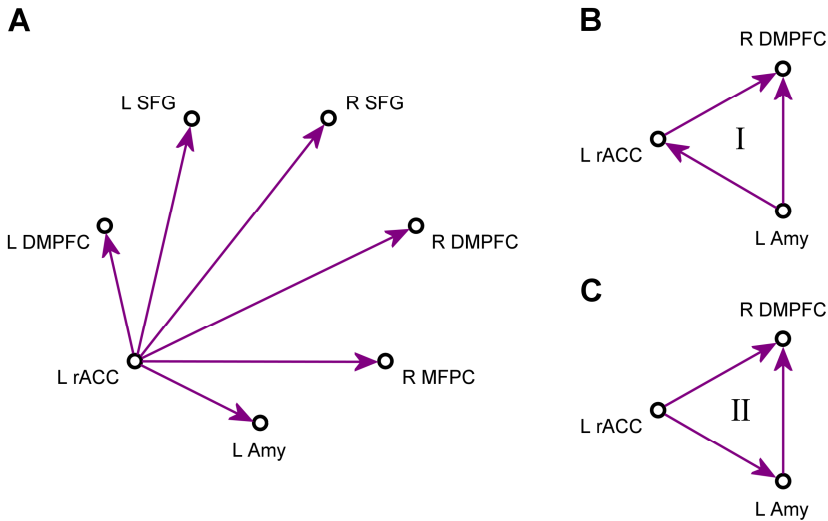

**Figure 5. Schematics of structural models used in the multivariate SVAR analyses.** (A) An example of a star model for instantaneous effects. A model of this kind was defined for each of the seven ROIs and examined in the multivariate structural vector autoregression (SVAR) analysis. (B,C) Two models for instantaneous effects, Model I and Model II, that provided the best $\chi^2$ fits to the experimental group data in the SVAR analyses for the system of three ROIs. A total of 24 structural models were optimized and compared for the system consisting of the left amygdala, the left rACC, and the right DMPFC (see text for details).

*SVAR Analysis*

Figure 5A exhibits a schematic of a star model for instantaneous effects of the left rACC on the other six regions. The matrix $A_0$ in Eq (1) had six free parameters in this case, and all the other matrix elements were set to zero. Average values of the path coefficients (mean±sem) for both instantaneous and lagged effects of the left rACC are shown in Fig. 6A. The corresponding linear trend statistics across the six experimental runs are included in Table 3. According to these data, the effects of the left rACC appeared very similar for four regions within the network: the left DMPFC, the right DMPFC, the right MFPC, and the left SFG. For these regions, the instantaneous effects for EG (denoted as $EG_0$, magenta) showed significant linear trends across the neurofeedback runs and across the entire experiment (Table 3, column $EG_0$). The lagged effects for EG (denoted as $EG_1$, red) exhibited no significant linear trends (Table 3, column $EG_1$), and the corresponding path coefficients were close to zero (Fig. 6A). The instantaneous effects for CG (denoted as $CG_0$, cyan) showed no significant linear trends (Table 3, column $CG_0$). However, the lagged effects for CG (denoted as $CG_1$, blue) exhibited negative trends, which were significant for the left DMPFC, the right DMPFC, the right MFPC, and the right SFG (Table 3, column $CG_1$).

The SVAR results describing the effects of the left rACC on the LA in Fig. 6A and Table 3 are inconclusive, however. The instantaneous effects for EG exhibited no significant trend (rACC=>LA: LT(RE…R3), $t(13)=$ 0.327, $P<0.748$; LT(RE…TR), $t(13)=0.455$, $P<0.657$). The lagged effects for EG showed a positive trend, which, however, was not significant (rACC=>LA: LT(RE…R3), $t(13)=$ 1.554, $P<0.144$; LT(RE…TR), $t(13)=$ 1.350, $P<0.200$).

Results of the multivariate SVAR analysis with a star model for instantaneous effects of the LA on the other six regions are shown in Fig. 6B, and the corresponding linear trend statistics are reported in Table 3. Only effects of the LA on three other regions are included. According to these data, the instantaneous effects of the LA on the left DMPFC and the right DMPFC for EG exhibited significant positive linear trends across the experimental runs (Table 3, column $EG_0$), while all the other LA effects did not show any significant trends.

The significant linear trends for the instantaneous effects in Fig. 6A and Fig. 6B (Table 3, column $EG_0$) generally survived FDR correction for multiple comparisons within the corresponding star models with six instantaneous interactions (rACC=>left DMPFC: $P<0.011$, $q<0.022$; rACC=>right DMPFC: $P<0.007$, $q<0.021$; rACC=>right MFPC: $P<0.003$, $q<0.018$; rACC=>left SFG: $P<0.034$, $q<0.051$; LA=>left DMPFC: $P<0.018$, $q<0.054$; LA=>right DMPFC: $P<0.010$, $q<0.054$).

Similar multivariate SVAR analyses were performed using star models for instantaneous effects of the other network regions. No significant linear trends emerged for the effects of the right MFPC, the left SFG, and the right SFG for EG. The instantaneous effects of the left DMPFC on the left rACC and the LA for EG showed positive linear trends that were either significant or trended toward significance (left DMPFC=>rACC: LT(RE…TR), $t(13)=$ 1.956, $P<0.072$; left DMPFC=>LA: LT(RE…TR), $t(13)=$ 2.380, $P<0.033$). Similarly, the instantaneous effects of the right DMPFC on the same two regions exhibited positive linear trends that trended toward significance (right DMPFC=>rACC: LT(RE…TR), $t(13)=2.126$, $P<0.053$; right DMPFC=>LA: LT(RE…TR), $t(13)=2.008$, $P<0.066$). Notably, these linear trends were less significant than those in the effects of the left rACC and the LA on the left DMPFC and the right DMPFC (Table 3, column $EG_0$).

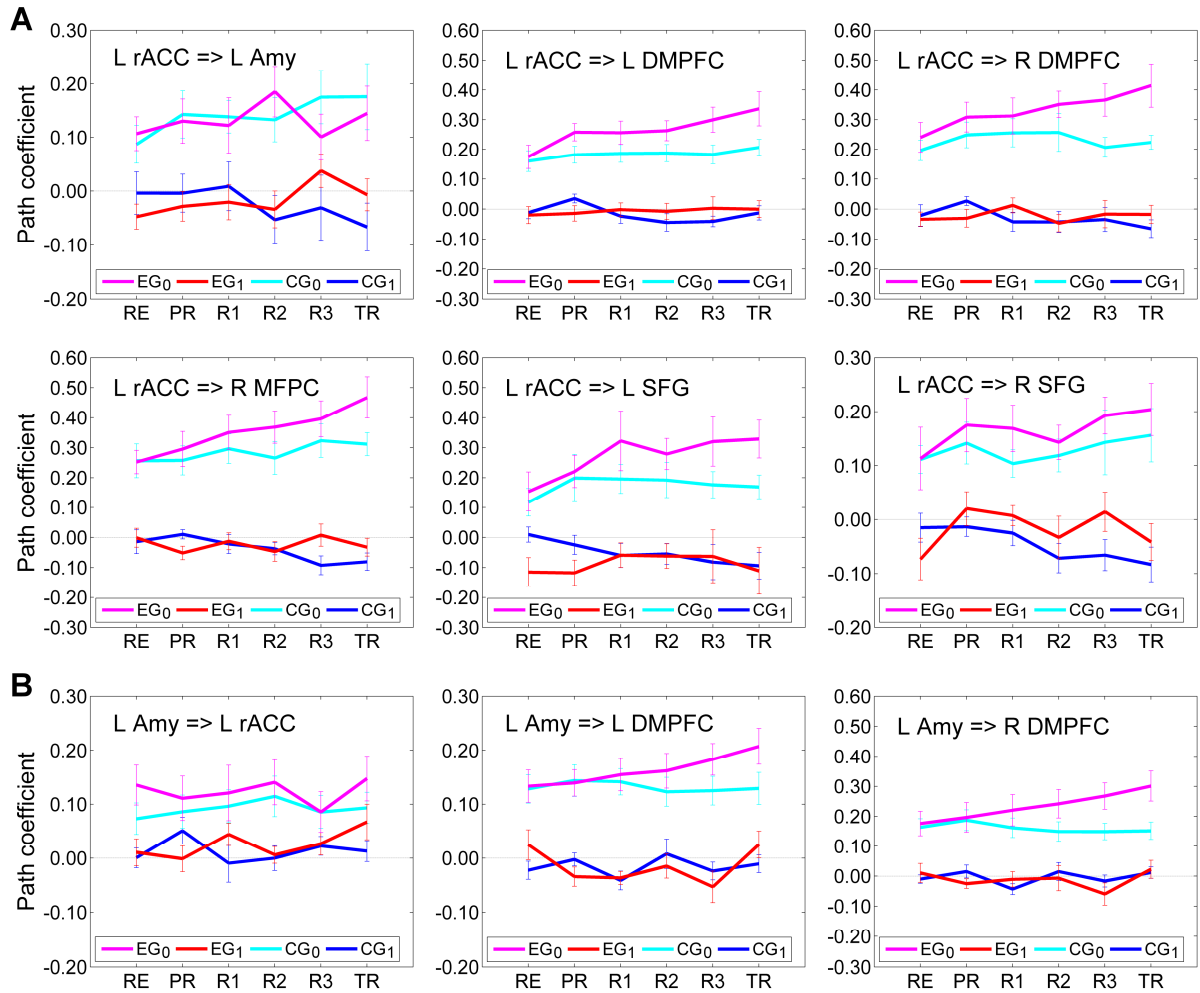

**Figure 6. Interactions suggested by the multivariate SVAR analyses for seven ROIs.** (A) Results of the multivariate first-order structural vector autoregression (SVAR) analysis for the network of seven ROIs using a star model for instantaneous effects of the left rACC (Fig. 5A). (B) Results of a similar SVAR analysis using a star model for instantaneous effects of the left amygdala. For the experimental group (EG), average path coefficients (mean±sem) for the instantaneous effects are depicted in magenta and denoted $EG_0$, and those for the lagged effects are depicted in red and denoted $EG_1$. For the control group (CG), average path coefficients for the instantaneous effects are shown in cyan and denoted $CG_0$, and those for the lagged effects are shown in blue and denoted $CG_1$.

Based on the results of the seven SVAR analyses with star models, we selected three regions that showed the most significant progressive changes in their instantaneous interactions across the experimental runs: the left rACC, the LA, and the right DMPFC (see *Discussion* for a detailed justification of this region selection). Further SVAR analyses were applied to this system of three ROIs. Overall, 24 SVAR models with $n=3$ were defined and optimized as described above (see *Data Analysis*). Among the 24 examined structural models, two models provided the most accurate descriptions of the instantaneous effects for EG. They are denoted as Model I and Model II, and depicted schematically in Fig. 5B and Fig. 5C, respectively. For Model I, the average $\chi^2$ value (mean±std) over three neurofeedback runs (Run 1, Run 2, Run 3) for 14 subjects in EG was $\chi^2=0.005\pm0.016$. For Model II, this average value was $\chi^2=0.006\pm0.020$. These values with $df=2$ correspond to $P=0.9975$ and $P=0.9970$, respectively, indicating that both models provided excellent fits to the instantaneous effects in the experimental time series data for the three ROIs.

Figure 7 exhibits average values of the path coefficients (mean±sem) in Model I and Model II. The instantaneous effects common to both Model I and Model II are shown in Fig. 7A, while the effects specific to each of the two models are shown in Fig. 7B and Fig. 7C, respectively. The corresponding linear trend statistics across the neurofeedback runs (RE…R3) are included in Table 4. According to these data, the instantaneous effects of the left rACC and the LA on the right DMPFC for EG exhibited significant positive linear trends (rACC=>R DMPFC: LT(RE…R3), $t(13)=3.053$, $P<0.009$; LA=>R DMPFC: LT(RE…R3), $t(13)=3.618$, $P<0.003$) (Table 4, column $EG_0$). The instantaneous interactions between the left rACC and the LA in both Model I and Model II did not show significant trends (Table 4, column $EG_0$). Importantly, the positive linear trend in the lagged effect of the left rACC on the LA in Model I for EG was significant (rACC=>LA: LT(RE…R3), $t(13)=2.422$, $P<0.031$) (Table 4, column $EG_1$). The same lagged effect in Model II showed a positive linear trend that was marginally significant (rACC=>LA: LT(RE…R3), $t(13)=1.776$, $P<0.099$). However, Model II also demonstrated a competition between the instantaneous and lagged effects of the left rACC on the LA for EG: increases in the instantaneous effect were accompanied by decreases in the lagged effect, and vice versa (R2, R3, TR in Fig. 7C, top). When the path coefficients for the instantaneous and lagged effects in Model II were summed for each subject and each run, the group results showed significant positive linear trends both across the neurofeedback runs and across the entire experiment (rACC=>LA: LT(RE…R3), $t(13)=2.615$, $P<0.021$; LT(RE…TR), $t(13)=2.823$, $P<0.014$). This means that the cumulative effect (i.e. $EG_0+EG_1$) of the left rACC on the LA in Model II showed a significant positive linear trend across the rtfMRI-nf training runs.

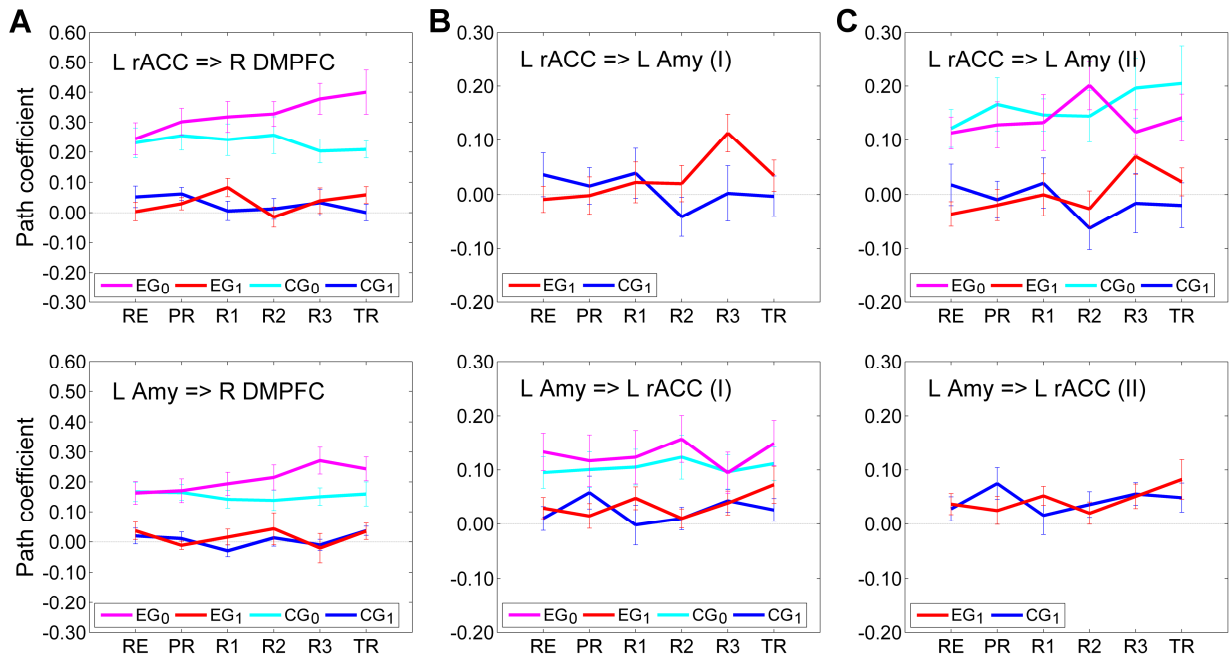

**Figure 7. Interactions suggested by the multivariate SVAR analyses for three ROIs.** Results of the multivariate SVAR analyses for the system of three ROIs – the left rACC, the left amygdala, and the right DMPFC – with the models for instantaneous effects depicted in Fig. 5 B,C. (A) Effects that are common to both Model I (Fig. 5B) and Model II (Fig. 5C). (B) Interactions between the left rACC and the left amygdala in the SVAR analysis with Model I (Fig. 5B). (C) Interactions between the left rACC and the left amygdala in the SVAR analysis with Model II (Fig. 5C). Notations are the same as in Fig. 6.

## Discussion

In this study, we applied structural vector autoregression (SVAR) modeling [33] to explore effective connectivity specific to the rtfMRI-nf training of emotional self-regulation [22]. This is the first application of SVAR analysis to rtfMRI-nf data to our knowledge.

The analysis of percent changes in BOLD signal activity, illustrated in Fig. 2 for three representative ROIs, shows similar activity patterns across the experimental runs for all seven ROIs in Fig. 1. The group BOLD activity results for the LA in Fig. 2 exhibit three distinct properties (see *ROI Analysis* above). *First*, a significant positive linear trend is observed for EG across the neurofeedback training runs (with the Rest run as the starting point) and across the entire experiment. This indicates that the mean LA BOLD activity increased progressively during the rtfMRI-nf training. *Second*, no significant difference between the mean BOLD activity levels for the last neurofeedback training run (Run 3) and the Transfer run is observed for EG. This demonstrates that the participants' learned ability to activate the LA during the rtfMRI-nf training generalized to the situation when the neurofeedback was no longer provided. *Third*, a significant difference in the mean BOLD activity levels is observed between EG and CG groups for Run 3 and for the Transfer run. This indicates that the ability to increase the LA activity was specific to EG. The left rACC and the right DMPFC ROIs exhibited BOLD activity properties that appear similar to those for the LA ROI (Fig. 2) on the basis of group results showing trends toward statistical significance (see *ROI Analysis*). These three properties reflect the important features of the experimental design, as discussed in [22]. In particular, the positive linear trend in the LA BOLD activity across the rtfMRI-nf training runs for EG arose due to the fact that the target level for the rtfMRI-nf bar was raised in a linear fashion from run to run (see *Subjects and Procedure*). The mean BOLD activity levels for the left rACC and the right DMPFC also showed linear trends for EG, but not for CG (Fig. 2). In general, persistence of a linear trend through the Transfer run depends on the extent to which a certain training effect generalizes beyond the actual training. The results of the VAR and SVAR network analyses were tested for such linear trends and group differences as signatures of rtfMRI-nf training effects.

The multivariate VAR analysis (Figs. 3 and 4, Tables 1 and 2) suggests that the left rACC plays a prominent role during the rtfMRI-nf training of the amygdala. According to the analysis, the left rACC exerted significant directional effects on the LA and the right MFPC during the last neurofeedback run (Run 3) for EG (Fig. 3A), but not for CG (Fig. 3B). Moreover, the effect of the left rACC on the LA was significantly enhanced during Run 3 compared to the Rest run for EG (Fig. 3C). Similarly, the effects of the left rACC on the other six regions during Run 3 either were significantly stronger or trended toward being stronger for EG than for CG (Fig. 3D). No significant differences were observed for the left rACC effects between the VAR results for Run 3 and for the Transfer run, suggesting that these effects persisted beyond the actual neurofeedback training. Furthermore, the left rACC effects on the other regions exhibited positive linear trends for EG (Fig. 4, Table 2), that were either significant or trended toward significance across experimental runs (see *VAR Analysis*). Taken together, the VAR results indicate that a positive dynamic functional coupling exists between the left rACC and the LA during the rtfMRI-nf training with positive emotion induction, and that this coupling is enhanced as the training progresses.

The VAR analysis also suggests that the right MFPC is actively engaged during the rtfMRI-nf procedure with real neurofeedback (EG, Fig. 3A), while the left SFG plays an active role during the procedure with sham neurofeedback (CG, Fig. 3B). These regions perform higher cognitive functions. The medial frontopolar cortex (BA 10) shows increased BOLD activity during mentalizing, i.e. attending to one's own emotions and mental states, as well as during multi-task coordination [44]. Both functions are recruited during the rtfMRI-nf training, particularly for EG. The medial superior frontal gyrus (BA 6) is involved in selection of action and task switching [45]. While both the ACC and the SFG are generally involved in decision-making, the ACC (unlike the SFG) has a fundamental role in relating actions to their consequences [45]. The active engagement of the left SFG instead of the left rACC during the rtfMRI-nf procedure for CG is likely a reflection of the fact that the sham neurofeedback provides information inconsistent with performance of the emotion induction task.

The series of the multivariate SVAR analyses, in which instantaneous effects of each network region on the other six regions were modeled independently, as illustrated in Fig. 5A, generally confirmed the important role of the left rACC. However, these analyses also demonstrated that the effects of the left rACC on the prefrontal regions are more accurately described by the instantaneous, than by the lagged effect terms (Fig. 6A, Table 3). The instantaneous effects of the left rACC on four prefrontal regions – the left DMPFC, the right DMPFC, the right MFPC, and the left SFG – exhibited significant positive linear trends across the experimental runs for EG, but not for CG (Fig. 6A, Table 3). The lagged effects corresponding to the same interactions were negligible in comparison, and did not exhibit positive trends (Fig. 6A, Table 3). Interestingly, the lagged effects of the left rACC for CG showed negative linear trends, that were significant for several regions (Fig. 6A, Table 3). The instantaneous effects of the LA on the left and right DMPFC exhibited significant positive linear trends for EG, while the corresponding lagged effects were negligible, and did not show significant trends for either EG or CG (Fig. 6B, Table 3). Notably, effects of the LA on the DMPFC regions did not emerge in the VAR analysis at all (Fig. 3), suggesting that such effects were much closer to instantaneous than to lagged ones (for $TR$=2 s). It should be noted that the SVAR analyses for the seven ROIs were completely independent of the VAR analysis, and were not informed by it in any way.

To enable a more accurate modeling of instantaneous effects and examine interactions between the left rACC and the LA with improved statistical power, we selected the system of three ROIs for further analyses: the left rACC, the LA, and the right DMPFC. This region selection was based on the following considerations. *First*, the instantaneous effects of both the left rACC and the LA on the bilateral DMPFC showed significant enhancements during the rtfMRI-nf training for EG in the SVAR analyses with the seven ROIs (Fig. 6, Table 3). In contrast, the instantaneous effects of the right MFPC and the bilateral SFG on the other regions did not exhibit any significant trends. *Second*, the rACC (BA 24) and the DMPFC (BA 9) have extensive direct anatomical connections with the amygdala, while connections of the MFPC (BA 10) and SFG (BA 6) with the amygdala are very scarce [46,47]. *Third*, the rACC and the DMPFC consistently show functional co-activation with the amygdala in various emotional tasks [48]. *Fourth*, the effects of the left rACC and the LA were similar for both the left and right DMPFC (Fig. 6, Table 3). The two DMPFC ROIs belong to the same functional area, and their centers are only 10 mm apart. Thus, it is sufficient to consider only one of the two regions. We chose the right DMPFC, because it experienced stronger instantaneous effects from both the left rACC and the LA (Fig. 6). Selection of the left DMPFC instead of the right DMPFC produced similar statistical results.

The SVAR analyses for the three selected ROIs indicated that the instantaneous effects within the system could be quite accurately described by two structural models (see *SVAR Analysis*). Model I (Fig. 5B, Fig. 7B) included the instantaneous effect LA=>rACC, while Model II (Fig. 5C, Fig. 7C) included the instantaneous effect rACC=>LA instead. The two interactions cannot be modeled simultaneously in SEM, because the two paths are non-recursive. The instantaneous effects of the left rACC and the LA on the right DMPFC were the same in both models, and showed significant positive linear trends across the neurofeedback runs (Fig. 7A, Table 4). The fact that both models provided similar-quality fits to the experimental data, as demonstrated by their $\chi^2$ values (see *SVAR Analysis*) suggests that the instantaneous interaction between the left rACC and the LA is bidirectional: rACC<=>LA. However, neither of the two instantaneous effects exhibited a significant linear trend across the neurofeedback runs (Fig. 7B,C, Table 4). In contrast, the lagged effect rACC=>LA exhibited a positive linear trend that was significant in Model I ($P<0.031$) and marginally significant in Model II ($P<0.099$). The latter result reflects the competition between the instantaneous and lagged rACC=>LA effects in Model II (Fig. 7C); their cumulative effect, nevertheless, showed a significant positive linear trend ($P<0.021$). We conclude that the rtfMRI-nf training targeting the LA [22] leads to a significant enhancement in the lagged effect of the left rACC on the LA. The instantaneous effects of both the left rACC and the LA on the bilateral DMPFC are also significantly enhanced.

SVAR network modeling [33] and a similar method called unified SEM [49] have been used for analysis of effective connectivity in neuroimaging data before (e.g. [50,51]). However, only VAR (Granger causality) [38] modeling has been previously applied to rtfMRI-nf data [30,52]. In particular, the authors of [30] demonstrated that patients with schizophrenia could learn to self-regulate their anterior insula BOLD activity using recall of emotionally relevant past experiences and rtfMRI-nf. A Granger causality analysis of the rtfMRI data suggested that effective connections among insula cortex, amygdala, and MPFC were enhanced at the end of the training [30]. While this conclusion is generally consistent with the results of the VAR analysis in the present work (Figs. 3,4), the analysis procedure of [30] differed from ours in several respects: i) selection of ROIs in [30] was based on fMRI activation data (taken partly from literature, partly from the actual GLM activation analysis), rather than on fMRI functional connectivity data; ii) a dorsal ACC ROI was included in the network in [30] rather than a rostral ACC ROI; iii) group causality maps were presented in [30] for two sessions with the strongest and the weakest insula regulation, but no statistical difference map was shown, and no statistical tests comparing results for the two sessions were reported. These methodological differences preclude a detailed comparison of the VAR results between the two studies.

Comparison of the SVAR and VAR results in the present work demonstrates that SVAR network modeling is clearly preferable to VAR if the fMRI repetition time is relatively long (*TR*=2 s in this study). In the described VAR analysis, the interactions with relatively short delay times (as suggested by the SVAR analyses) either appeared as lagged effects, or did not appear at all (Figs. 3,4 vs Fig. 6). At the same time, the SVAR analyses for three ROIs (Fig. 7, Table 4) confirmed the VAR result indicating that the lagged effect of the left rACC on the LA increased progressively during the rtfMRI-nf training (Fig. 4, Table 2). This lagged effect could, in principle, be interpreted as a Granger causality between the activities of neuronal populations. However, possible group-level differences in the hemodynamic response functions for the left rACC and the LA in the present study are unknown. It should be noted that Granger causal inferences are quite robust with respect to inter-regional hemodynamic response differences, provided that the temporal resolution is sufficiently high and fMRI noise is sufficiently low [53,54]. Unlike both VAR and SVAR, the DCM method [20] can explicitly account for hemodynamic response variability, but this would require measurements of the hemodynamic response functions for each ROI in every subject. It has been suggested that the two approaches – Granger causality (VAR) and DCM – may be converging, and that Granger causality may potentially provide candidate models for DCM [55]. Irrespective of a modeling method, an effective connectivity analysis of rtfMRI-nf data would benefit from a higher temporal resolution, which can be achieved, for example, by using SENSE [37] with higher acceleration factors ($R=2$ in this work). Because rtfMRI-nf training is usually accompanied by increased head motion, a more efficient correction of motion artifacts in fMRI time series (e.g. [56]) would also improve reliability and accuracy of modeling results.

The effective connectivity analysis, reported in this work, suggests that the left rACC plays an important role during the rtfMRI-nf training [22], modulating the left amygdala and other regions of the examined network. This conclusion is consistent with the results of multiple neuroimaging studies that have highlighted the role of left rACC in emotion regulation [8,13-19,21]. A recent study of self-regulation of emotion networks using rtfMRI-nf with positive mood induction also showed an increase in the left rACC activity as a result of the rtfMRI-nf training [27]. The SVAR analyses for three ROIs (Fig. 7, Table 4) demonstrate top-down modulation of the amygdala by the left rACC, similar to the one revealed by the bivariate DCM analysis in [17]. However, the dynamic functional coupling between the two regions is positive in our study, i.e. activity of the left rACC leads to an increased activity of the left amygdala. While many previous studies examining emotion control/regulation focused on the PFC-amygdala coupling when participants engaged in *down-regulation* of *negative emotions*, the current study focused on *up-regulation* of *positive emotions*. This procedural difference explains the positive rACC-amygdala coupling found here. The prominent role of the left rACC suggests that the rtfMRI-nf procedure [22] may engage subprocesses of automatic ("implicit") emotion regulation in addition to voluntary ("explicit") emotion regulation.

In contrast to the active role of the left rACC, the bilateral DMPFC experienced the directional effects of both the left rACC and the LA, according to the SVAR analyses (Figs. 6, 7). The DMPFC has consistently shown co-activation with the amygdala in a variety of emotional tasks [48]. Furthermore, the DMPFC is the only frontal region that exhibits co-activation with brainstem limbic structures, such as the hypothalamus and the periaqueductal gray matter, thought to be critical for physiological effects of emotion [48]. The enhancement in the instantaneous LA=>DMPFC effect during the rtfMRI-nf procedure may be important for practical applications of neuromodulation, because electrophysiological activity of the DMPFC can be measured by scalp EEG and used to provide EEG neurofeedback [31].

Our results are also consistent with animal studies that have directly explored anatomical connections and neuronal interactions between the PFC and the amygdala. For example, a study of the laminar distribution of connections between the PFC and the amygdala using injections of neural tracers in rhesus monkeys [46] showed that the ACC areas BA 24 and 25 (along with the posterior OFC) had the densest connections with the amygdala. Moreover, these ACC areas issued more projections to the amygdala than they received, suggesting a similar pattern for the flow of information [46]. In contrast, the DMPFC (BA 9) had more input connections from the amygdala than output connections to the amygdala [46]. Our SVAR modeling results, showing the enhancements in the rACC=>LA and LA=>DMPFC effects associated with rtfMRI-nf, are consistent with these neuroanatomical properties. An electrical microstimulation study in cats [57] demonstrated that stimulated neuronal firing in the mPFC was associated with an increased firing probability for neurons in the basolateral amygdala with typical time lags of 20-40 ms. This result suggested the existence of excitatory projections of the mPFC to the basolateral amygdala [57]. Similarly, a microstimulation study in rats [58] showed that the prelimbic subregion of the mPFC, involved in control over emotional-cognitive aspects of behavior, had excitatory projections to the basolateral amygdala.

Our results suggest that the rtfMRI-nf approach affords a powerful non-invasive tool for i) targeting selected brain regions and modulating their BOLD activity, ii) identifying and modulating activity of other brain regions engaged due to the rtfMRI-nf procedure, and iii) exploring the resulting network interactions. In particular, our results point to the rACC as a promising target for rtfMRI-nf training of emotion regulation along with the amygdala. Self-regulation of the rACC using rtfMRI-nf may be relevant for investigation and treatment of mood and anxiety disorders, particularly posttraumatic stress disorder (PTSD). Patients with PTSD show hypo-activation of the rACC (together with the dorsal ACC, the ventromedial PFC, and the thalamus), when responding to negative emotional stimuli (versus neutral or positive stimuli), compared to healthy participants [59,60]. This abnormal hypoactivation is associated with hyper-activation of the amygdala, and correlates with PTSD symptom severity [60]. Thus, an abnormally attenuated functional coupling is observed between the rACC and the amygdala in patients with PTSD compared to healthy subjects [59]. The effective connectivity analysis, reported in this work, indicates that the rtfMRI-nf training targeting the left amygdala [22] leads to a progressive enhancement in positive functional coupling and directional influence of the left rACC on the amygdala for positive emotion induction. This result suggests a novel therapeutic approach for reducing severity of PTSD symptoms, in which both the rACC and the amygdala are simultaneously used as target regions for rtfMRI-nf training.

**Competing interest**

**Acknowledgments**

This work was supported by the Laureate Institute for Brain Research and the William K. Warren Foundation. The funders had no role in study design, data collection and analysis, decision to publish, or preparation of the manuscript. We would like to thank Dr. Gang Chen of the National Institute of Mental Health for his helpful advices regarding SVAR modeling.

**Table 1. Effects of the rACC on the other six network regions according to the multivariate VAR analysis.** The table contains meta-analytic group values of the VAR path coefficients ($\alpha$) with the corresponding FDR $q$-values (in square brackets), as well as group differences in the path coefficients ($\Delta\alpha$) with the corresponding uncorrected $P$-values. The four data columns (A,B,C,D) correspond to the four subplots (A,B,C,D) in Fig. 3.

|  | **A** | **B** | **C** | **D** |
|---|---|---|---|---|
| Effect | $\alpha$ [$q$] | $\alpha$ [$q$] | $\Delta\alpha$ [$P$] | $\Delta\alpha$ [$P$] |
| L rACC => L Amy | 0.0857 [0.024]* | −0.0271 [0.767] | 0.1099 [0.010]* | 0.1102 [0.077] |
| L rACC => L DMPFC | 0.0888 [0.061] | −0.0004 [0.986] | 0.0588 [0.266] | 0.0833 [0.030]* |
| L rACC => R DMPFC | 0.0899 [0.141] | −0.0138 [0.776] | 0.0591 [0.361] | 0.0984 [0.090] |
| L rACC => R MFPC | 0.1298 [0.026]* | −0.0355 [0.681] | 0.0725 [0.279] | 0.1651 [0.006]* |
| L rACC => L SFG | 0.0798 [0.289] | −0.0191 [0.718] | 0.1374 [0.057] | 0.1080 [0.027]* |
| L rACC => R SFG | 0.0695 [0.178] | −0.0207 [0.700] | 0.1119 [0.039]* | 0.0908 [0.015]* |

*indicate significant effect

**Table 2. Trends in the rACC effects on the other six network regions according to the multivariate VAR analysis.** The table contains linear trend $t$-statistics for the VAR path coefficients across the neurofeedback runs (RE…R3) with the corresponding group $P$-values. Each data row corresponds to a subplot in Fig. 4. Notations are the same as in Fig. 4.

|  | **EG** | **CG** |
|---|---|---|
| Effect | $t(13)$ [$P$] | $t(13)$ [$P$] |
| L rACC => L Amy | 2.348 [0.035]* | −0.912 [0.379] |
| L rACC => L DMPFC | 1.961 [0.072] | −1.692 [0.115] |
| L rACC => R DMPFC | 1.437 [0.174] | −0.619 [0.547] |
| L rACC => R MFPC | 2.513 [0.026]* | −1.645 [0.124] |
| L rACC => L SFG | 2.921 [0.012]* | −1.136 [0.276] |
| L rACC => R SFG | 2.330 [0.037]* | −1.552 [0.145] |

*indicate significant effect

**Table 3. Trends in the rACC and amygdala effects on the other network regions according to the multivariate SVAR analyses for seven ROIs.** The table contains linear trend $t$-statistics for the SVAR path coefficients across the six experimental runs (RE…TR) with the corresponding group $P$-values. Each data row corresponds to a subplot in Fig. 6. Notations are the same as in Fig. 6.

|  | **EG$_0$** | **EG$_1$** | **CG$_0$** | **CG$_1$** |
|---|---|---|---|---|
| Effect | $t(13)$ [$P$] | $t(13)$ [$P$] | $t(13)$ [$P$] | $t(13)$ [$P$] |
| L rACC => L Amy | 0.455 [0.657] | 1.350 [0.200] | 1.501 [0.157] | −1.581 [0.138] |
| L rACC => L DMPFC | 2.967 [0.011]* | 0.742 [0.471] | 1.807 [0.094] | −2.280 [0.040]* |
| L rACC => R DMPFC | 3.188 [0.007]* | 0.357 [0.727] | 0.051 [0.960] | −2.646 [0.020]* |
| L rACC => R MFPC | 3.697 [0.003]* | −0.077 [0.940] | 1.475 [0.164] | −3.628 [0.003]* |
| L rACC => L SFG | 2.364 [0.034]* | 0.410 [0.688] | 0.548 [0.593] | −1.826 [0.091] |
| L rACC => R SFG | 1.156 [0.269] | 0.614 [0.550] | 0.760 [0.461] | −2.398 [0.032]* |
| L Amy => L rACC | 0.016 [0.988] | 1.230 [0.241] | 0.763 [0.459] | −0.085 [0.934] |
| L Amy => L DMPFC | 2.720 [0.018]* | −0.162 [0.874] | −0.458 [0.654] | 0.492 [0.631] |
| L Amy => R DMPFC | 3.034 [0.010]* | −0.160 [0.876] | −1.082 [0.299] | 0.559 [0.586] |

*indicate significant effect

**Table 4. Trends in the rACC and amygdala effects according to the multivariate SVAR analyses for three ROIs.** The table contains linear trend *t*-statistics for the SVAR path coefficients across the neurofeedback runs (RE…R3) with the corresponding group *P*-values. Each data row corresponds to a subplot in Fig. 7. An empty cell means an absence of the corresponding interaction from a model. Notations are the same as in Fig. 7.

| Effect (Model) | $EG_0$ $t(13)$ [$P$] | $EG_1$ $t(13)$ [$P$] | $CG_0$ $t(13)$ [$P$] | $CG_1$ $t(13)$ [$P$] |
|---|---|---|---|---|
| L rACC => R DMPFC (I,II) | 3.053 [0.009]* | 0.286 [0.780] | −0.674 [0.512] | −1.151 [0.270] |
| L Amy => R DMPFC (I,II) | 3.618 [0.003]* | −0.508 [0.620] | −1.004 [0.334] | −1.126 [0.281] |
| L rACC => L Amy (I) |  | 2.422 [0.031]* |  | −1.147 [0.272] |
| L Amy => L rACC (I) | −0.407 [0.690] | 0.172 [0.866] | 0.313 [0.759] | 0.303 [0.767] |
| L rACC => L Amy (II) | 0.636 [0.536] | 1.776 [0.099] | 0.919 [0.375] | −1.044 [0.316] |
| L Amy => L rACC (II) |  | 0.282 [0.783] |  | 0.279 [0.785] |

*indicate significant effect